\documentstyle[12pt,epsf]{article}
\begin{document}
\def\lsim{\mathrel{\lower4pt\hbox{$\sim$}}\hskip-12pt\raise1.6pt\hbox{$<$}\;}
\def\BAR{\bar}
\def\fm{{\cal M}}
\def\fl{{\cal L}}
\def\n{\delta}
\def\gsim{\mathrel{\lower4pt\hbox{$\sim$}}
\hskip-10pt\raise1.6pt\hbox{$>$}\;}

\vspace*{-.5in}
\rightline{AMES-HET 99-03}   
\rightline{UCRHEP-T251}
\rightline{BNL-HET-99/11}
\begin{center}

{\large\bf 
Graviton Production By Two Photon Processes 
In Kaluza-Klein Theories With Large Extra Dimensions\\
}

\vspace{.3in}

David Atwood$^{1}$\\ 
\noindent Department of Physics and Astronomy, Iowa State University, Ames,
IA\ \ \hspace*{6pt}50011\\
\medskip

Shaouly Bar-Shalom$^{2}$\\ 
\noindent Department of Physics, University of California, Riverside, 
CA 92521\\
\medskip
and\\
\medskip

Amarjit Soni$^{3}$\\
\noindent Theory Group, Brookhaven National Laboratory, Upton, NY\ \ 
11973\\
\footnotetext[1]{email: atwood@iastate.edu}
\footnotetext[2]{email: shaouly@phyun0.ucr.edu}
\footnotetext[3]{email: soni@bnl.gov}
\end{center}
\vspace{.2in}

\begin{quote}
{\bf Abstract:}  

We consider the production of gravitons via two photon fusion in
Kaluza-Klein theories which allow $TeV$ scale gravitational interactions.
We find that the processes $\ell^+\ell^-\to \ell^+\ell^-+graviton$, with
$\ell=e$, $\mu$ can put quite stringent bounds on such theories. For
example, with two extra dimensions at the Next Linear Collider with a
center of mass energy of 500 (1000) $GeV$ attainable bounds on the scale
of the gravitational interactions can reach about 6 (9) $TeV$. 

\end{quote} 
\vspace{.3in} 
\newpage

Gravity is the weakest force of nature and,
although it ultimately controls the shape of the entire universe, its role
in fundamental interactions remains obscure. This is due to the fact that
gravity remains weak until the unreachably high scale of the Planck mass
and thus there is no experimental data to construct a theory of gravity at
small distances.

Of course the lack of experimental evidence has not deterred the
construction of theories to account for the properties of gravitation at
short distances. In this Letter we will consider certain Kaluza-Klein
theories which contain additional compact dimensions besides the
four space-time dimensions.

In such theories it was traditionally assumed that the compact dimensions
form a manifold which is unobservably small (perhaps at the Planck scale)
and thus remain hidden. However, recent advances in
M-theory~\cite{mtheory}, a Kaluza-Klein theory in which there are 11 total
dimensions suggest another possible scenario~\cite{add1}.  In models
proposed in~\cite{add1,add2}, $\delta$ of these extra dimensions may be
relatively large while the remaining dimensions are small. In this class
of theories, the known fermions, the strong, weak and electromagnetic
forces exist on a 4-brane while gravity may act in $4+\delta$ dimensions.
The size of these extra dimensions, $R$, is related to an effective Planck
mass, $M_D$, according to~\cite{add1}:

\begin{eqnarray}
8\pi R^\delta M_D^{2+\delta}\sim M_P^2
\label{Rsize}
\end{eqnarray}

\noindent
where $M_P=1/\sqrt{G_N}$ is the Planck mass and
$G_N$ is Newton's constant.
Indeed the effective Planck mass at which gravitational effects become
important may be as small as $O(1~TeV)$ in which case such effects
may be probed in collider experiments.

In this scenario, at distances $d<R$ the Newtonian inverse square law will
fail~\cite{add1}. If $\delta=1$ and $M_D$=$1~TeV$, then R is of the order
of $10^{8}~km$, large on the scale of the solar system, which is clearly
ruled out by astronomical observations.  However, if $\delta\geq 2$ then
$R< 1~mm$; there are no experimental constraints on the behavior of
gravitation at such scales~\cite{cavin} so these models are possible.

Astonishingly enough if $M_D\sim 1~TeV$ then gravitons may be readily
produced in accelerator experiments.  This is because the extra dimensions
give an increased phase space for graviton radiation. Another way of
looking at this situation is to interpret gravitons which move parallel to
the 4 dimensions of space time as the usual gravitons giving rise to
Newtonian gravity while the gravitons with momentum components
perpendicular to the brane are effectively a continuum of massive objects.
The density of gravitons states is given by~\cite{add1,add2,wells,taohan}:

\begin{eqnarray}
D(m^2)={dN\over d m^2}={1\over 2} S_{\delta-1} 
{\BAR M_P^2 m^{\delta-2}\over M_D^{\delta+2}}
\end{eqnarray}

\noindent
where $m$ is the mass of the graviton, $\BAR M_P=M_P/\sqrt{8\pi}$
and 
$S_k=2\pi^{(k+1)/2}/\Gamma[(k+1)/2]$.
The probability of graviton emission  may thus become large when the sum
over the huge number of graviton modes is considered. 

Gravitons with polarizations that lie entirely within the physical
dimensions are effective spin 2 objects which we consider in this Letter. 
Gravitons with polarizations partially or completely perpendicular to the
physical brane are vector and scalar objects which we will not consider
here since they couple more weakly than the spin 2 type.

The compelling idea that gravity may interact strongly at $TeV$ scale
energies has recently led to a lot of phenomenological activity. $TeV$
scale gravity can be manifested either directly through real graviton
production, leading to a missing energy signal, or indirectly through
virtual graviton exchanges. Thus, existing and future high energy
colliders can place bounds on the scale and the number of extra dimensions
in these theories by looking for such 
signals~\cite{hepph9811337},~\cite{wells}~--~\cite{newrizzo}.

Typically, direct signals drop as $(E/M_D)^{\delta+2}$, where $E$ is the
maximum energy carried by the emitted gravitons. Therefore, the best
limits on $M_D$ from the existing experimental data at LEPII, Tevatron and
HERA are obtained for the case $\n=2$.  For example, existing LEPII data on
$\sigma(e^+e^- \to \gamma + missing~energy)$ already places the bound,
$M_D \gsim 1~TeV$ for $\n=2$ via the process $e^+e^- \to \gamma +G$ (see
references~\cite{hepph9811337, wells, hepph9903294}). For $\n=4$ the
limit is $M_D \gsim 700$ GeV. A NLC with c.m. energy $\gsim 1~TeV$ can
push this limit up to $M_D \gsim 6~TeV$ (for $\n=2$) and $M_D \gsim 4~TeV$
(for $\n=4$) \cite{hepph9811337}. In hadronic colliders, the signal $p \bar
p \to jet + missing~energy$ can proceed by the subprocesses $q \bar q \to
g G$, $q (\bar q) g \to q (\bar q) G$ and $gg \to gG$.  Using these, the
existing Tevatron data on $\sigma(p \bar p \to jet + missing~energy)$
places the limits $M_D \gsim 750$ GeV for $\n=2$ and $M_D \gsim 600$ GeV
for $\n=4$, while the LHC will be able to probe $M_D$ up to $\sim 7~TeV$ for
$\n=2$ \cite{hepph9811337,wells}.

The present bounds obtained from indirect signals associated with virtual
graviton exchanges are typically $M_D \gsim 500 -700$ GeV via processes
such as $e^+e^- \to \gamma \gamma,~ZZ,~W^+W^-$ (LEPII)
\cite{hepph9902263}, $e^+q \to e^+q$ or $e^+g \to e^+g$ (HERA)
\cite{hepph9812486,hepph9901209}, $p \bar p \to t \bar t + X$ (Tevatron)
\cite{hepph9811501}, and $M_D \gsim 1~TeV$ via processes such as $q \bar
q,gg \to \ell^+\ell^-$ (Tevatron) and $e^+e^- \to f \bar f$ (LEPII)
\cite{hepph9811356,hepph9901209}. Future colliders such as the NLC and the
LHC will be able to push these limits to several $TeV$'s through the study
of these signals. Clearly other new physics can also give rise to similar
signals so they may be used to bound $TeV$ scale 
gravitation theories but can only confirm them with more extensive 
analysis, for example by angular distributions of 
final state particles (e.g.~\cite{hepph9811356}). It should also be noted 
that the predictions in virtual graviton processes 
have some uncertainties since they depend on the sum over the
Kaluza Klein (KK) tower of the massive  excitations which is not
fully determined without knowing the full quantum gravity theory.

In this paper we investigate another possible direct signal of strongly
coupled low energy gravity via the process $e^+e^- \to e^+e^- G$ ($G$=spin
2 graviton) which proceeds predominantly through the t-channel $\gamma
\gamma$ (or $ZZ$) fusion subprocesses $\gamma \gamma (ZZ) \to G$. Since
these photons tend to be collinear, the process $e^+e^- \to e^+e^- G$ is
significantly enhanced compared to $s$-channel processes. We find that the
resulting signal is robust and possibly more useful for detecting or
constraining some low energy gravity scenarios at the energy scales of a
future NLC.

Let us now consider the excitation of spin 2 graviton modes through
photon-photon and $ZZ$ fusion. Such a process could be probed at an
$e^+e^-$ collider where the effective photon luminosity is generated by
collinear photon emission. The complete process is therefore $e^+e^-\to
e^+ e^- G$ through the diagram shown in Fig.~1.  In principle the other
diagrams where the graviton is attached to the fermion lines or directly
to the gauge-fermion vertex will also contribute, but the process in
Fig.~1 should be dominant due to the enhancement of collinear gauge boson
emission.

Let us consider first the case of photon-photon fusion.  The cross section
of this process may be estimated through the Weiszacker-Williams leading
log approximation~\cite{peskin_schroeder}.  Thus if $\sum|\fm(\hat s)|^2$
is the matrix element for $\gamma\gamma\to G$, where $G$ is a graviton of
mass $m=\sqrt{\hat s}$, then in this approximation the total cross 
section for $e^+e^-\to e^+e^- G$ is given by:

\begin{eqnarray}
\sigma(e^+e^-\to e^+e^-G)
={\pi\eta^2\over 4 s}
\int_0^1 
{f(\omega)\over \omega} 
D(\omega s)
\sum|\fm(\omega s)|^2 d\omega
\label{wwaprox}
\end{eqnarray}

\noindent
where $s$ is the center of mass energy of the collision,
\begin{eqnarray}
f(\omega)=\left [ (2+\omega)^2\log(1/\omega)
-2(1-\omega)(3+\omega)\right ] /\omega ~~~ {\rm and} ~~~
\eta=\alpha \log\left [s/(4m_e^2)\right ] /(2\pi).
\nonumber
\end{eqnarray}

Using the effective Lagrangian for the $G\gamma\gamma$ coupling derived
in~\cite{wells,taohan}, we obtain: 

\begin{eqnarray}
\sum|\fm(\hat s)|^2 
=2 {\hat s^2\over \BAR M_P^2}
\label{matel}
\end{eqnarray}

\noindent
Note that the explicit dependence on $\BAR M_P$ will cancel when 
multiplied by the density of graviton states.
This is typical of reactions involving real graviton emission.
We therefore obtain the total cross section in this approximation:

\begin{eqnarray}
\sigma_{\gamma\gamma}(e^+e^-\to e^+e^-G)
=
{\alpha^2\over 16\pi s}
S_{\delta-1} 
\left [{\sqrt s\over M_D}\right ]^{\delta+2}
F_{\delta\over 2} 
\log^2\left [ {s\over 4 m_e^2}\right ]
\end{eqnarray}

\noindent
where $F_k=\int_0^1 f(\omega)\omega^k d\omega$.

In Fig.~2, the solid curves give the total cross section as a function of
$s$ given $M_D=1~TeV$ for $e^+e^-\to e^+e^-G$ in the cases where
$\delta=$2 and 6 (corresponding to the upper and lower solid curves) while
the thin dashed curve is the cross section for $\mu^+\mu^-\to \mu^+\mu^-G$
with $\delta=2$ which would be applicable to a muon collider.

Experimental considerations suggest that perhaps the full cross section
which is given in the above is not observable.  Gravitons couple very
weakly to normal matter and thus a radiated graviton will not be detected
in the detector.  Therefore, the signature for the reaction would be

\begin{eqnarray}
e^+e^-\to e^+e^-~+~missing~mass. 
\nonumber
\end{eqnarray}

\noindent Since this cross section is dominated by emission of photons at
a small angle, the outgoing electrons will therefore also be deflected by
a small angle. Although one can expect that the electrons will suffer an
energy loss, a significant portion of the electrons will not be deflected
out of the area of the beam pipe and so may not be directly detected.  To
obtain a more realistic estimate one must therefore select events where
the electron is deflected enough to be detected. Moreover, there is a
Standard Model background to this signal from the process $e^+e^-\to
e^+e^-\nu_\ell\bar\nu_\ell$. The component of this cross section which
results from $ZZ$ fusion, $ZZ\to \nu_\ell\bar\nu_\ell$, in particular, has
a $P_T$ distribution similar to the signals we consider. We calculate this
background using the effective boson approximation~\cite{evba}.  This
background is $0.25~fb$ for $\sqrt{s}=500~GeV$ and $1.6~fb$ at
$\sqrt{s}=1~TeV$.  Let us now consider three possible methods for
detection of this signal.

First, one could take advantage of the fact that a significant amount of
energy present in the initial collision is lost to the unobservable
graviton.  In Fig.~3, the normalized missing mass distribution is shown as
a function of $\omega=\hat s/s$ where $\hat s$ is the missing mass squared
of the graviton.  In this approximation, this distribution is not changed
by the value of $\sqrt{s}$, $M_D$ or any systematic cut imposed on the
transverse momentum $P_T$ of the outgoing electrons.  The distribution is
shown for $\delta=2$ (solid), $\delta=4$ (dashed), $\delta=6$ (dotted) and
$\delta=8$ (dot-dash). In principle it might be possible to separate the
reduced energy electrons form the outgoing electrons of the collision at a
$e^+e^-$ collider through downstream dipole magnets but the large
bremsstrahlung radiation generated by the disruption of the collision
probably makes such an electron difficult or impossible to detect. At a
muon collider, perhaps a Roman Pot could find reduced energy muons which
were deflected from the main beam however the decay electrons in the muon
collider environment may make this difficult also. Clearly experimental
innovations are required to detect the full cross section and we will not
consider this further.

Secondly, if both of the electrons are given enough of a transverse
momentum that they may be detected in the detector or the end-caps, events
of the desired type may be identified. Using the leading log
approximation, one can use Eqn.~\ref{wwaprox} with $\eta$ replaced by
$\hat \eta(P_{Tmin})=\alpha \log\left [s/(4P_{Tmin})\right ] /(2\pi)$
where $P_{Tmin}$ is the minimum transverse momentum of the outgoing
electron which is accepted.  If one imposes this cut on the two outgoing
electrons one obtains the cross section as a function of $\hat s$ shown in
Fig.~2 with the dotted curve for the case of $P_{Tmin}=10$~GeV with
$M_D=1~TeV$ and $\delta=2$, while the heavy dot-dot-dash curve is for
$\delta=4$.  These curves would be the same at both electron and muon
colliders since the transverse momentum cut is well above the lepton mass.
The missing mass spectra under this cut should also correspond to the
curves shown in Fig.~3.

The missing mass spectrum for the background discussed above is shown in
the case of $\sqrt{s}=1~TeV$ with the dot-dot-dash line (The normalization
of this curve is reduced by 1/10 for clairity). Clearly this distribution
differs markedly from the signal and cuts may thus be used to enhance the
discrimination between signal and background. To obtain bound on $M_D$ we
will consider a cut of $\omega>0.16$ which reduces the background to
$0.19~fb$. In contrast, the signal is reduced by a factor of 0.42 in the
case of $\delta=2$, 0.82 in the case of $\delta=4$, 0.96 in the case of
$\delta=6$ and 0.99 in the case of $\delta=8$.

Thirdly, one could identify events where only one of the electrons has a
transverse momentum greater than $P_{Tmin}$.  This would in effect be
replacing $\eta^2$ in Eqn.~\ref{wwaprox} with
$\eta_{eff}^2=2\eta(\eta-\eta(P_{Tmin}))$. The resultant cross sections
are shown in Fig.~2 with the dot-dash curve for $P_{Tmin}=20$~Gev.  In
this case, the energy of the detected electron will be markedly reduced
from the beam energy since the graviton mass distribution increases at
high masses.  In Fig.~4 we show the normalized missing energy ($E_{miss}$)
spectrum as a function of $x=2E_{miss}/\sqrt{s}=E_{miss}/E_{beam}$ for the
detected electron where $\delta=2$ (solid), $\delta=4$ (dashed),
$\delta=6$ (dotted) and $\delta=8$ (dot-dash). In this leading log
approximation, the curves of Fig.~4 are largely independent of $P_{Tmin}$.

The missing energy spectrum for the background is shown in the case of
$\sqrt{s}=1~TeV$ with the dot-dot-dash line (with 1/2 the normalization). 
Again a cut in $x$ can enhance the the signal with respect to the
background somewhat. If we impose the cut $x>0.2$, the background is
reduced to $0.78~fb$, while the signal is reduced by a factor of 0.72 in
the case of $\delta=2$, 0.93 in the case of $\delta=4$, 0.99 in the case
of $\delta=6$ and 0.997 in the case of $\delta=8$.

Let us now consider the related process $e^+e^- \to ZZ e^+ e^- \to e^+ e^-
G$ which can likewise be estimated by the effective vector boson leading
log approximation.  In general the cross section is given by a sum over
cross sections for $ZZ\to G$ in various helicity combinations together
with the helicity dependent structure functions given in ~\cite{evba}.
Here there is considerable simplification since in this approximation
where the boson momenta are taken collinear with their parent leptons, the
only amplitude which contributes are the cases where the bosons are
transverse and of opposite helicities. As with the photon, we use the
effective Lagrangian from~\cite{wells} and obtain the cross section in
this approximation:

\begin{eqnarray}
\sigma_{ZZ}(e^+e^-\to e^+e^-G)
=
{y^2\alpha^2\over 16\pi s}
S_{\delta-1} 
\left( {\sqrt s\over M_D} \right )^{\delta+2} 
\left [ F_{\delta\over 2}^Z(s)+z^2 H_{\delta\over 2}^Z(s) \right ]
\log^2 \left( {s\over M_Z^2} \right )
\end{eqnarray}

\noindent
where

\begin{eqnarray}
&&x_w=\sin^2\theta_w,\ \ \ \ 
y={1-4x_w+8x_w^2\over 8 x_w (1-x_w)}, \ \ \ \
z={1-4x_w\over 2(1-4x_w+8x_w^2)},
\nonumber\\
&&F_k^Z(s)=\int_{4 m_Z^2\over s}^1 \omega^k f(\omega) d\omega
\nonumber\\
&&H_k^Z(s)=-\int_{4 m_Z^2\over s}^1 4 \omega^{k}
\left[(4+\omega)log(\frac{1}{\omega}) - 4(1-\omega) \right]
\end{eqnarray}

\noindent 
and $f(\omega)$ is defined as for the case of photons.

In Fig.~2 the thick dashed curve shows the total cross section for this
process given $M_D=1~TeV$ and $\delta=2$. This cross section is flat in
$P_T$ for $P_T<O(m_Z)$ and therefore $O(10~GeV)$ cuts in $P_T$ of the
outgoing leptons will not reduce this greatly.  For the same reason the
cross section at a $\mu\mu$ collider will be the same.

In Table~1 we consider the limits that may be placed on theories with
extra dimensions using these $e^+e^-\to e^+e^-G$ processes. We consider
three possible accelerator scenarios: 
$\sqrt{s}=200~GeV$ and a total 
integrated luminosity of $2.5~fb^{-1}$ (for LEP-200);
$\sqrt{s}=500~GeV$ and a total integrated luminosity of $50~fb^{-1}$; 
$\sqrt{s}=1~TeV$ and a total integrated luminosity of 
$200~fb^{-1}$. These last two cases correspond to a future NLC.
For
$\sqrt{s}=1~TeV$ we will impose the cut of 
$\omega>0.16$ on both the signal and the background in the 
case where both electrons are subject to the  $P_{Tmin}=10~GeV$ cut
and a cut of $x>0.2$ if only one electron is subjected to this cut.
We define the lower limit on $M_D$ in each case to be the value which will
yield 10 events in each scenario 
or a signal of statistical significance 
of $3\sigma$ above the background. 
For $\delta=2$, $4$ and $6$ we 
consider detection either via the full cross section (if that were somehow
observable) or via the signal with the cut $P_{Tmin}=10~GeV$ on just one
outgoing electron or both outgoing electrons.

As can be seen, using the two electron signal, at the $200~GeV$ collider,
a limit of about 0.5-2~$TeV$ (depending on $\delta$) may be placed on $M_D$;
using the $500~GeV$ collider a limit of about 1-6~$TeV$ may be obtained and
with a $1~TeV$ collider a limit of about 2.5-9~$TeV$ may result. Clearly
the limit on $M_D$ decreases somewhat as $\delta$ increases. 
Obviously, with less stringent cuts
and/or using a single high $P_T$ lepton tag
the lower limit on $M_D$ may be 
increased somewhat.

If a signal is seen, the missing mass distributions in Fig.~3 and the 
missing energy distributions in Fig.~4 will help distinguish these 
theories from other new physics candidates and also help to determine how 
many extra dimensions are present. 

\bigskip

We are grateful to Jose Wudka for discussions.  One of us (DA) thanks the
UCR Theory Group for hospitality. This research was supported in part by
US DOE Contract Nos.  DE-FG01-94ER40817 (ISU), DE-FG03-94ER40837 (UCR) and
DE-AC02-98CH10886 (BNL)

\newpage

\begin{center}
{\Large\bf Table 1}
\end{center}
$$
\begin{tabular}{||c|c|c|c|c||} 
\hline
\multicolumn{5}{||c||}{$\delta=2$}\\ \hline & 
&
{No cut}&
{$P_{Tmin}=10~GeV$}&
{$P_{Tmin}=10~GeV$}\\ 
$\sqrt{s}$ & 
$\int {\cal L} dt$&
{\ }&
{(one electron)}&
{(two electrons)}\\ 
\hline
200~GeV&
2.5~$fb^{-1}$&
2.4~TeV&
1.8~TeV&
1.0~TeV
\\ \hline
500~GeV&
50~$fb^{-1}$&
8.2~TeV&
6.4~TeV&
4.4~TeV
\\ \hline
1000~GeV&
200~$fb^{-1}$&
14.4~TeV&
8.9~TeV&
6.2~TeV
\\ 
\hline
\hline
\multicolumn{5}{||c||}{$\delta=4$}\\ \hline
& 
&
{No Cut}&
{$P_{Tmin}=10~GeV$}&
{$P_{Tmin}=10~GeV$}\\ 
$\sqrt{s}$ & 
$\int {\cal L} dt$&
{\ }&
{(one electron)}&
{(two electrons)}\\ 
\hline
200~GeV&
2.5~$fb^{-1}$&
1.0~TeV&
0.8~TeV&
0.5~TeV
\\ \hline
500~GeV&
50~$fb^{-1}$&
3.0~TeV&
2.6~TeV&
2.0~TeV
\\ \hline
1000~GeV&
200~$fb^{-1}$&
5.4~TeV&
4.1~TeV&
3.5~TeV
\\ 
\hline
\hline
\multicolumn{5}{||c||}{$\delta=6$}\\ \hline
& 
&
{No Cut}&
{$P_{Tmin}=10~GeV$}&
{$P_{Tmin}=10~GeV$}\\ 
$\sqrt{s}$ & 
$\int {\cal L} dt$&
{\ }&
{(one electron)}&
{(two electrons)}\\ 
\hline
200~GeV&
2.5~$fb^{-1}$&
0.6~TeV&
0.5~TeV&
0.4~TeV
\\ \hline
500~GeV&
50~$fb^{-1}$&
1.8~TeV&
1.6~TeV&
1.3~TeV
\\ \hline
1000~GeV&
200~$fb^{-1}$&
3.3~TeV&
2.7~TeV&
2.4~TeV
\\ 
\hline
\end{tabular}
$$

\bigskip
\bigskip

{\bf Table 1:} The limits on the parameter $M_D$ are given for $\delta=2$,
4 and 6. In each case three accelerator scenarios are considered with
$\sqrt{s}=200~GeV$, $500~GeV$ and $1000~GeV$ with luminosities
$2.5~fb^{-1}$, $50~fb^{-1}$ and $200~fb^{-1}$ respectively. The signals
considered are based on the total cross section, the cross section with
one electron passing the $P_{Tmin}=10~GeV$ cut and the cross section with
both electrons passing the $P_{Tmin}=10~GeV$ cut. The limit which is
placed on $M_D$ is based on the criterion of 10 events for the given
luminosity or a significance of 3$\sigma$ above the background.

\newpage
\begin{center}
{\Large\bf Figure Captions}
\end{center}

\bigskip

\noindent {\bf Figure 1:} The dominant Feynman diagram for $e^+e^-\to e^+
e^-G$ through an effective photon or $Z$ sub-process.

\bigskip

\noindent {\bf Figure 2:} 
The cross sections for various processes are shown as a function of
$\sqrt{s}$. 
The solid lines are the total cross sections for $\delta=2$ (upper curve)
and $\delta=6$ (lower curve). 
The dotted line is for the case that both the outgoing electrons are subject 
to the cut $P_{Tmin}=10~GeV$ and for $\delta=2$. 
The dot-dash line 
is obtained again with
$\delta=2$ but now only  one of the outgoing electrons is subject to the cut
$P_{Tmin}=20~GeV$.
The thick dot-dot-dash line 
is for
$\delta=4$ where both of the outgoing electrons are subject to the
cut $P_{Tmin}=10~GeV$.
The thick dashed line shows the total cross section 
for $\delta=2$ via the $ZZ$ process. 
The dashed line gives the cross section for $\mu^+\mu^-\to\mu^+\mu^-G$
for $\delta=2$ via the $\gamma\gamma$ process. 
In all cases we take $M_D=1~TeV$.

\bigskip

\noindent {\bf Figure 3:} The normalized differential cross section as a
function of the scaled 
missing (graviton) 
invariant mass squared ($\omega=\hat s/s$) for
$\delta=2$ (solid line), $\delta=4$ (dashed line),
$\delta=6$ (dotted line), $\delta=8$ (dot-dash line).  These curves are
not greatly effected by the $P_{Tmin}$ cut, $M_D$ or $s$. 
The dot-dot-dash curve shows $(1/10)d\sigma/(\sigma d \omega)$ for the 
background.

\bigskip

\noindent {\bf Figure 4:} The normalized differential cross section as a
function of the missing energy of the single detected electron.
See also caption to Fig.~3.
Here, the dot-dot-dash curve shows $(1/2)d\sigma/(\sigma d x)$ for the 
background.

\newpage

\
\begin{figure}
\vspace*{0 in}
\hspace*{+1.0 in}
\epsfxsize 3.0 in
\mbox{\epsfbox{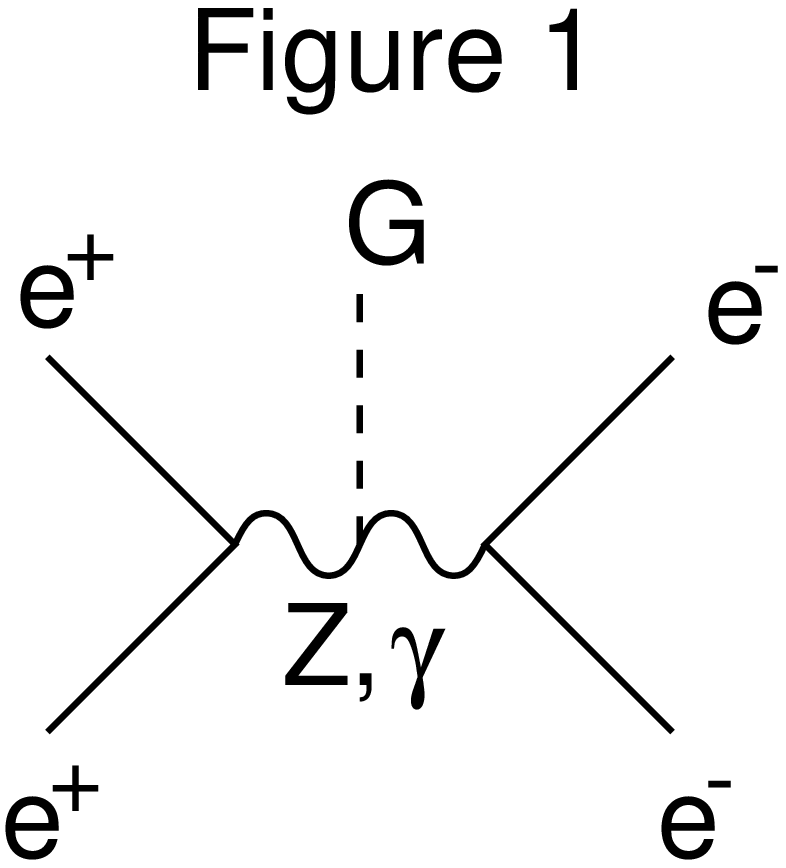}}
\end{figure}

\newpage

\
\begin{figure}
\vspace*{0 in}
\hspace*{-1.0 in}
\epsfxsize 7.0 in
\mbox{\epsfbox{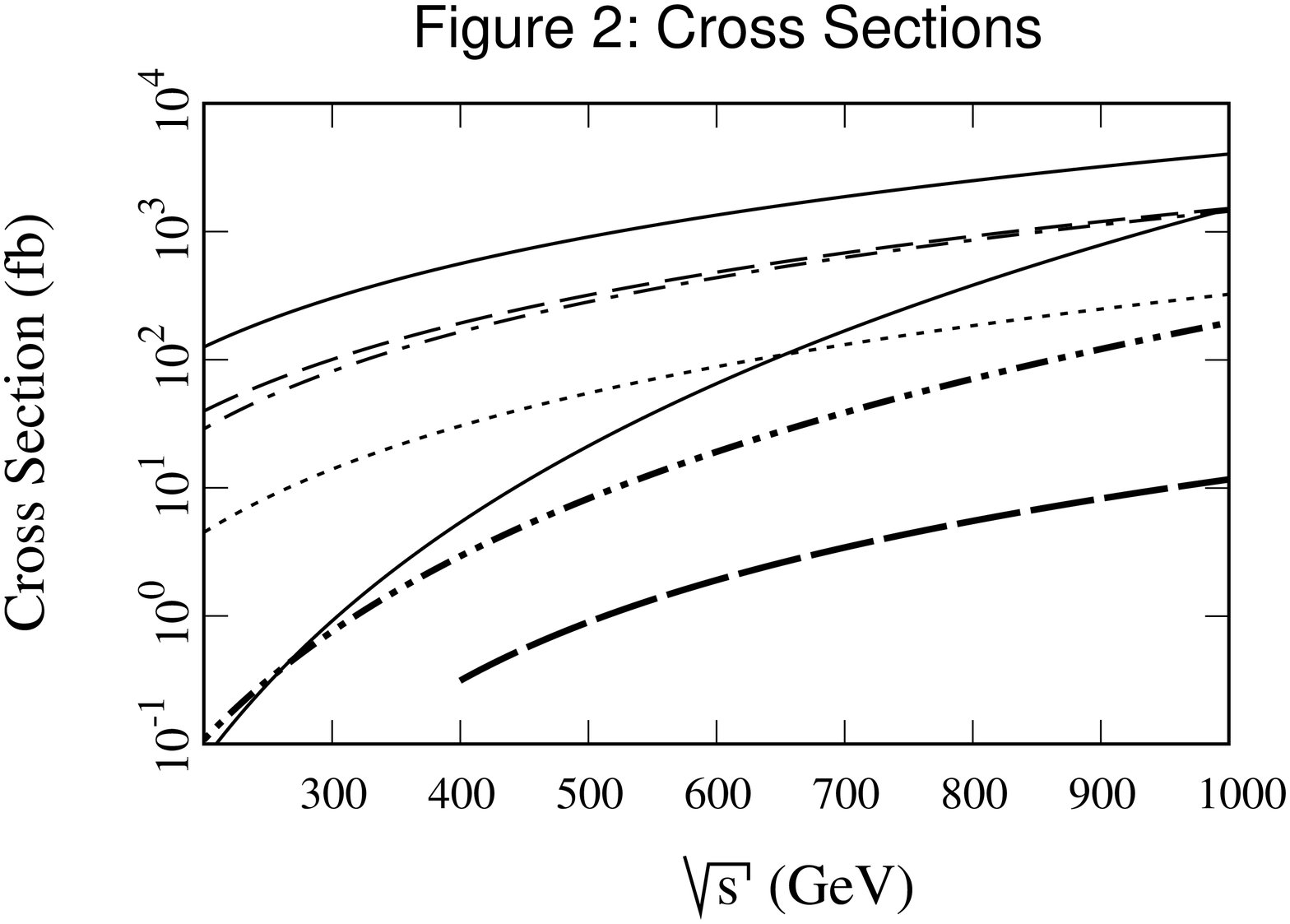}}
\end{figure}

\newpage

\
\begin{figure}
\vspace*{0 in}
\hspace*{-0.5 in}
\epsfxsize 6.0 in
\mbox{\epsfbox{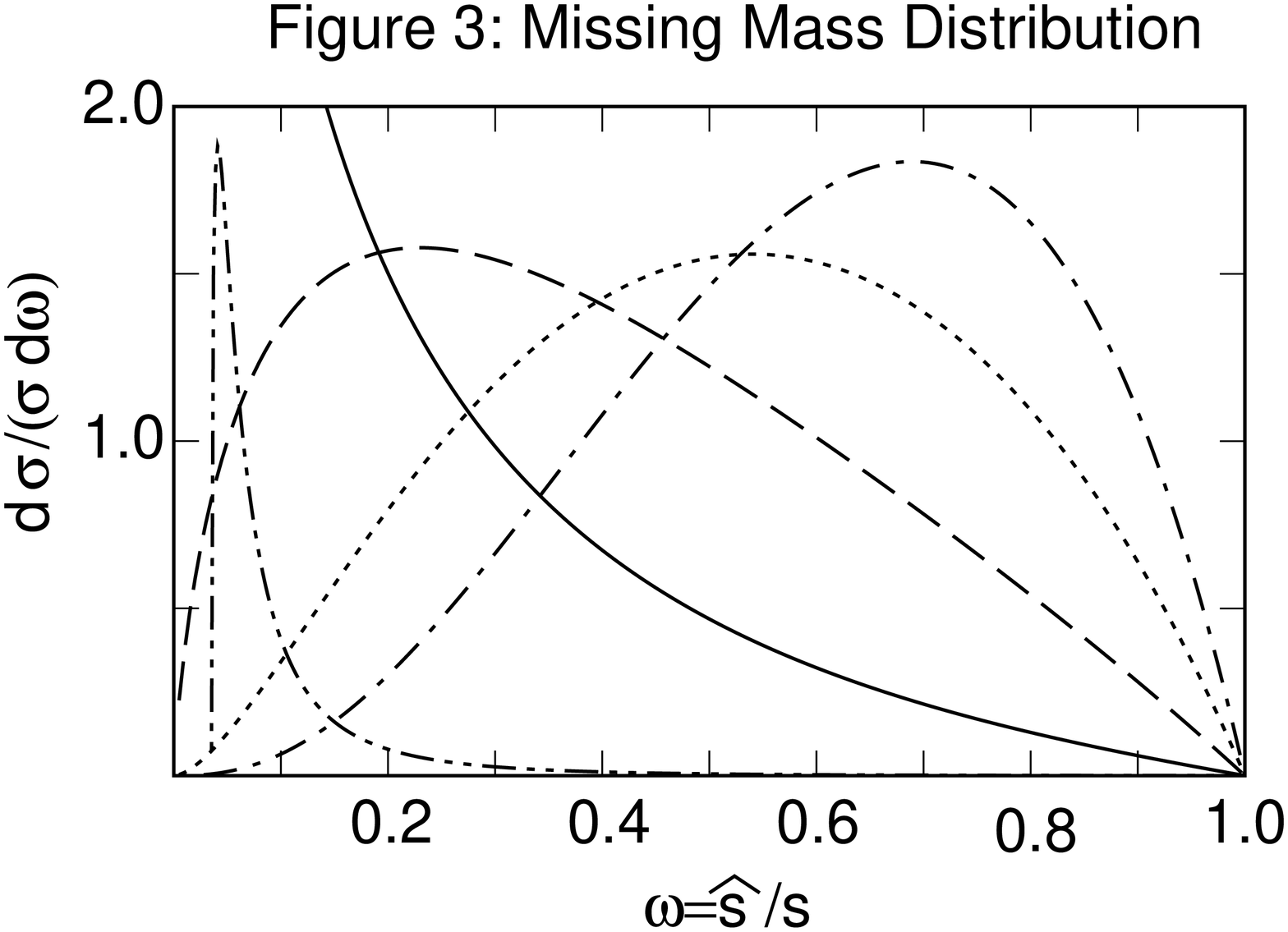}}
\end{figure}

\newpage

\
\begin{figure}
\vspace*{0 in}
\hspace*{-0.5 in}
\epsfxsize 6.5 in
\mbox{\epsfbox{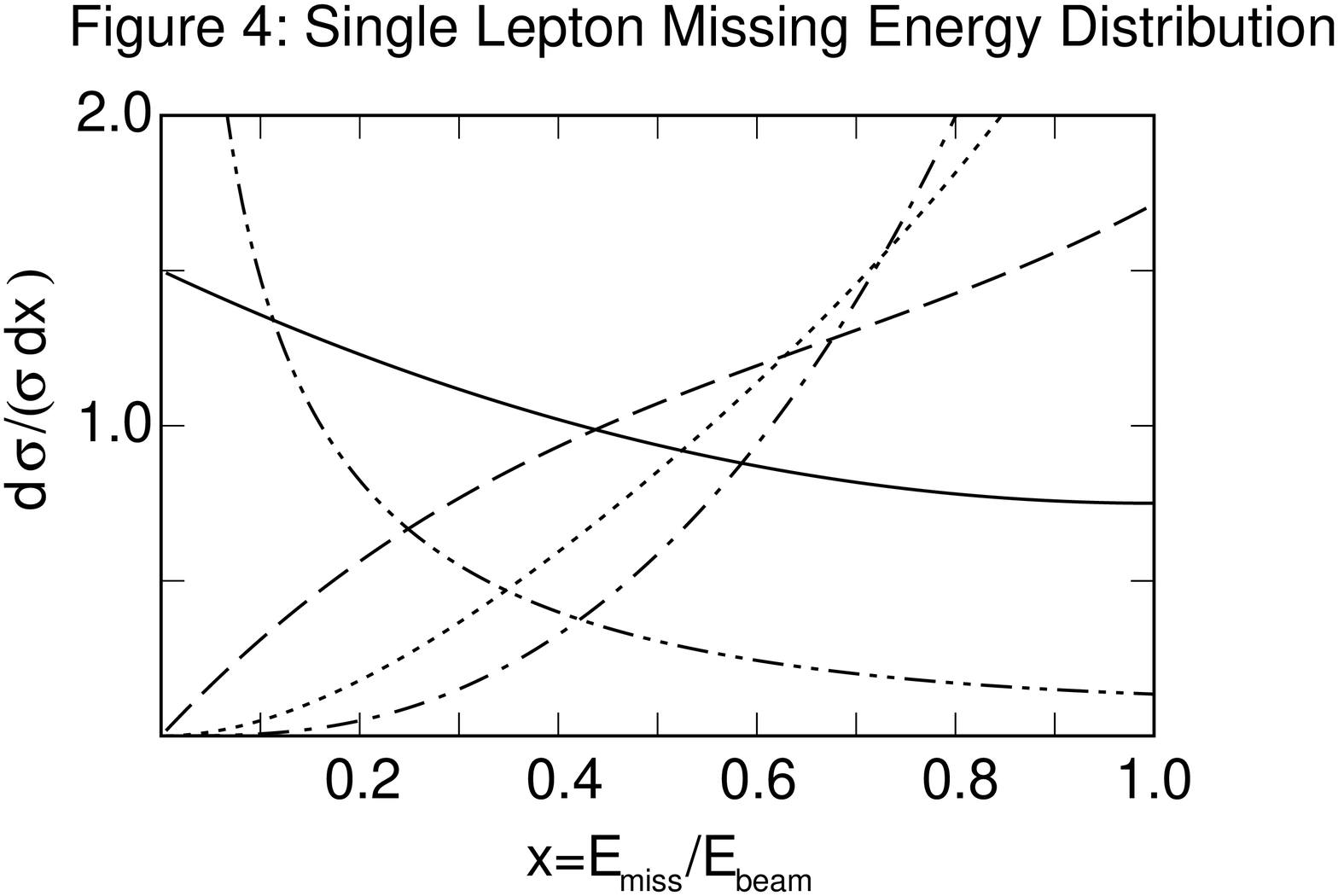}}
\end{figure}

\end{document}